\documentclass[twoside]{article}
\usepackage{tabulary,graphicx,times,caption,fancyhdr,amsfonts,amssymb,amsbsy,latexsym,amsmath}
\usepackage[utf8]{inputenc}
\pdfoutput=1
\usepackage{url,multirow,morefloats,floatflt,cancel,tfrupee,textcomp,colortbl,xcolor,pifont}
\usepackage[nointegrals]{wasysym}
\usepackage{graphics}
\usepackage{ulem}
\usepackage{multirow}
\usepackage{float}
\usepackage{inputenc}
\usepackage{xspace}
\usepackage{url}
\usepackage{epstopdf}
\usepackage{amsfonts}
\usepackage{amsmath}
\usepackage{amssymb}
\usepackage{extarrows}
\graphicspath{ {./images/} }
\allowdisplaybreaks

\makeatletter

\usepackage{ifxetex}
\ifxetex\else
  \usepackage{dblfloatfix}
\fi

\@ifundefined{subparagraph}{
\def\subparagraph{\@startsection{paragraph}{5}{2\parindent}{0ex plus 0.1ex minus 0.1ex}%
{0ex}{\normalfont\small\itshape}}%
}{}

\def\URL#1#2{\@ifundefined{href}{#2}{\href{#1}{#2}}}

\def\UrlOrds{\do\*\do\-\do\~\do\'\do\"\do\-}%
\g@addto@macro{\UrlBreaks}{\UrlOrds}

\makeatother

\usepackage[paperheight=11in,paperwidth=8.3in,margin=2.5cm,headsep=.7cm,top=2.5cm]{geometry}
\usepackage[T1]{fontenc}

\renewenvironment{abstract}
	{\trivlist\item[]\leftskip0pt\par\vskip4pt\noindent
  	\textbf{\abstractname}\mbox{\null}\\}
	{\par\noindent\endtrivlist}

\def\keywords#1{\par\medskip\par\noindent\textbf{Keywords}: #1\par}

 \date{} \emergencystretch 8pt

\makeatletter
\def\author#1{\gdef\@author{\hskip-\tabcolsep%
	\parbox{\textwidth}{\raggedright\bfseries#1\\[1pc]}}}
\def\address[#1]#2{\g@addto@macro\@author{\\\hskip-\tabcolsep\parbox{\textwidth}{\raggedright%
	\normalsize\normalfont\textsuperscript{#1}#2}}}
\let\addresslink\textsuperscript
\def\correspondence#1{\g@addto@macro\@author{\\\hskip-\tabcolsep\parbox{\textwidth}{\raggedright%
	\vspace*{10pt}\normalsize\normalfont~\\#1~\\[12pt]}}}
\def\email#1{\g@addto@macro\@author{\\\hskip-\tabcolsep\parbox{\textwidth}{\raggedright%
	\normalsize\normalfont Emails: #1}}}

\def\title#1{\gdef\@title{\vspace*{-30pt}%
	\raggedright\textbf{\@journaltitle}~\\%
  \raggedright\bfseries\ifx\@articleType\@empty\vspace*{20pt}\else%
  \vspace*{20pt}\@articleType\vspace*{20pt}\\\fi#1}}
\let\@journaltitle\@empty \def\journaltitle#1{\gdef\@journaltitle{{\normalfont\itshape#1}}}
\let\@articleType\@empty \def\articletype#1{\gdef\@articleType{{\normalfont\itshape#1}}}

\let\@runningHead\@empty \def\RunningHead#1{\gdef\@runningHead{{\normalfont #1}}}

\usepackage{fancyhdr}
\fancypagestyle{headings}{\fancyhf{}
  \fancyhead[R]{\itshape\@runningHead}
  \fancyfoot[C]{\thepage}}
\fancypagestyle{plain}{%
	\fancyhf{}\fancyhead[R]{Prepared for submission to American Journal of Modern Physics}
  \fancyfoot[C]{\thepage}}
\makeatother

\usepackage[%
	numbers,sort&compress%
  ]{natbib}

\usepackage{float,xcolor}

\journaltitle{Theoretical Physics}

\begin{document}

\title{Fermionic and bosonic partition functions at imaginary chemical potential as Bloch functions}

\author{
		Evangelos G. Filothodoros\addresslink{1}}		
\address[1]{Institute of Theoretical Physics, Aristotle University of Thessaloniki, Thessaloniki, Greece.}

\correspondence{Correspondence should be addressed to 
    	Evangelos G. Filothodoros; efilotho@physics.auth.gr}
    	
\correspondence{ORCID:https://orcid.org/0000-0002-5898-7288}



\maketitle 

\begin{abstract}
We point out that the phase transitions of the $d+1$ Gross-Neveu and $CP^{N-1}$ models at finite temperature and imaginary chemical potential can be mapped to transformations of Hubbard-like regular hexagonal to square lattice with the intermediate steps to be specific surfaces (irregular hexagonal kind) with an ordered construction based on the even indexed Bloch-Wigner-Ramakrishnan polylogarithm function. The zeros and extrema of the Clausen $Cl_d(\theta) $ function play an important role to the analysis since they allow us not only to study the fermionic and bosonic theories and their phase transitions but also the possibility to explore the existence of conductors arising from the correspondence between the partition functions of the two models and the Bloch and Wannier functions that play a crucial role in the tight-binding approximation in solid state physics. 

\keywords{Fermion-boson map; Hubbard lattice; conductor; Bloch function; Wannier function}
\end{abstract}
    
\section{1.\,\,Introduction}

In the realm of condensed matter physics
the Hubbard model stands as a cornerstone in the study of electronic correlations within quantum materials \cite{Kambis, Dong}. It describes electrons in a solid that interact with each other through short-range repulsive interactions. Specifically, the Hubbard model introduces a contact interaction between particles of opposite spin on each site of a lattice. When applied to electron systems, these interactions are typically expected to be repulsive, arising from the screened Coulomb interaction. In this work, we believe that if we examine in detail the setup of the higher dimensional thermal windows for fermions and bosons at imaginary chemical potential, we may connect them to the transformation of Hubbard-like lattices to square lattice and the appearance of insulator's identities. An Appendix contains some technical details and useful formulae for the Bloch-Wigner function.


\section{2.\,\,The Hubbard model and a brief review of the charges of the $U(N)$ fermionic Gross-Neveu and the bosonic CP$^{N-1}$  models at imaginary chemical potential in odd dimensions}

In the absence of a chemical potential, the fermionic Gross-Neveu and bosonic CP$^{N-1}$ models exhibit distinct symmetry-breaking patterns at finite temperature $T$. The Gross-Neveu model displays a parity-broken phase at low temperatures, which vanishes beyond a critical temperature. Conversely, the CP$^{N-1}$ model follows the usual continuous symmetry-breaking pattern at zero temperature, but this broken phase ceases to exist for $T>0$. Instead, when the coupling reaches its critical value at $T=0$, a finite-temperature scaling regime emerges, characterized by a non-zero thermal mass for the scalars.
From a previous work, detailed in \cite{Filothodoros:2016txa}, reveals that introducing an imaginary chemical potential alters this situation. The phase structures of both models can be mapped onto each other. Additionally, we have explored the significance of the celebrated Bloch-Wigner function on the unit circle, leveraging it in our calculations of phase transition-thermal windows \cite{Filothodoros:2018}. In a pursuit to illuminate bosonization physics, we have delved into the fermion-boson map in higher dimensions, comparing our continuum models’ phase transformations with those of specific Hubbard-like lattices. By utilizing the zeros and maximizations of Bloch-Wigner functions introduced by Zagier \cite{Filothodoros:2016txa, Filothodoros:2018, Zagier1, Zagier2}, we have uncovered an alternative perspective that draws parallels to the emergence of conductors. We consider that the charge arising as the eigenvalue of the charge $N$-normalized fermion number density operator can separate the insulating from the conducting state of a Hubbard-like model of fermions \cite{Murugan:2016zal}. These investigations extend across arbitrary odd dimensions.

\subsection{\textit{2.1.\,\,The Hubbard model at imaginary chemical potential}}

The Hubbard model is an approximate model that offers an insight view into how interacted electrons give as a result insulating or conducting phenomena in a crystal. Let's see the Hubbard Hamiltonian.

\begin{equation}
H=-t\sum_{[i,j]_s}c^{\dagger}_{is}c_{js}+U\sum_i n_{iu}n_{id}-i\mu\sum_i (n_{iu}+n_{id})
\end{equation}
where the notations $iu$ and $id$ imply spin up and spin down respectively, $n$ is the number operator and $c^\dagger, c$ are the creation and destruction operators that create an electron in a state $\Phi_R$ and destroy an electron in a state $\Phi_R$, where $\Phi_R$ is the Wannier function. Wannier functions are orthogonal functions, used in solid state physics, that refer to localized molecular orbitals with periodic boundary conditions. Since they are orthogonal 

\begin{equation}
\int_{crystal}\Phi(x)_1^*\Phi(x)_2dx=0
\end{equation}
or 
\begin{equation}
e^{\int_{crystal}\Phi(x)_1^*\Phi_2(x)dx}=1
\end{equation}
The first term is the kinetic term which shows how electrons move from one site to a neighbour site of the lattice. The second term adds an energy $U$ if the site already has two electrons.
At last we have the imaginary chemical potential term. So, what does the chemical potential actually do? Firstly, we use the imaginary chemical potential to avoid the sign problem that arises in Monte Carlo simulations (configurations in the partition function that may be negative or complex). Mainly, the imaginary chemical potential places the fermions and bosons in such positions, creating a complex structure of hexagonal lattices, sometimes creating an accumulation of charge and sometimes its nullification \cite{Langfeld}. So we have an image of a conductor for the first case and an insulator for the second case.

\subsection{\textit{2.2.\,\,The charge $Q_{df}$ of fermions}}
The GN model in $d$ Euclidean dimensions is described by the action \cite { Filothodoros:2018}
\begin{align}
 S_{1f} = -\int_0^\frac{1}{T} \!\!\!dx^0\int \!\!d^{d-1}\bar{x} \left[\bar{\psi }^{a}(\slash\!\!\!\partial  -i\gamma_0\alpha)\psi ^{a}
+\frac{G_d}{2({\rm Tr}\mathbb I_{d-1})N}\left (\bar{\psi }^{a}\psi ^{a}\right )^{2} +i\alpha NQ_{df}\right]\,,
\end{align}
with  $Q_{df}$ the $N$-normalized $d$-dimensional fermionic number density and $a=1,2,..N$.  For odd $d$ we take the dimension of the gamma matrices to be ${\rm Tr}{\mathbb I}_{d-1}=2^{\frac{d-1}{2}}$.
We see that the model also has $3$ terms. The first one is the kinetic term, the second is the interaction term that contributes to the energy of the fermions and the the third one is the imaginary chemical potential term with $Q$ the number eigenvalue of the fermion number operator.
The canonical partition function for an auxiliary scalar field $\sigma$ is expressed as follows:
\begin{align}
\label{GNPF1}
Z_{1f}(\frac{1}{T},Q_{df})&=\int({\cal D}\alpha)({\cal D}\sigma)e^{-N S_{1f,eff}}\,,\\
\label{GNPF2}
S_{1f,eff}&=iQ_{df}\int_0^\frac{1}{T} \!\!\!dx^0\!\!\int\!\!d^{d-1}\bar{x}\,\alpha -\frac{{\rm Tr}\mathbb I_d}{2G_d}\int_0^\frac{1}{T}\!\!\!dx^0\!\!\int d^{d-1}\bar{x}\,\sigma^2+\rm Tr\ln\left(\slash\!\!\!\partial-i\gamma_0\alpha+\sigma\right)_\frac{1}{T}\,.
\end{align}

where $\sigma$ plays the role of the mass of the fermion condensate.

The $d$-dimensional charge gap equation becomes:

\begin{equation}
\label{gap31}
iQ_{df}=\lim_{\epsilon\rightarrow 0}{\rm Tr}\mathbb I_{d-1}T\int^\Lambda\!\!\frac{d^{d-1} \bar{p}}{(2\pi)^{d-1}}
\sum_{n=-\infty}^\infty\frac{e^{i\omega_n\epsilon}(\omega_n-\alpha_*)}{\bar{p}^2+(\omega_n-\alpha_*)^2+\sigma_*^2},.
\end{equation}

We will discuss below in some detail the cases $d=5$ and $d=7$ in order to exhibit some of the general features of the higher dimensional models. Starting with $d=5$ and using the results of \cite{Filothodoros:2016txa, Filothodoros:2018}, we have the form of the charge gap equation as
\begin{equation}
\frac{\pi^2 Q_{5f}}{T^4}-3i\left[D_4(-z_*)+
\frac{1}{6}\ln^2\!|z_*|D_2(-z_*)\right]=0
\end{equation}
where $z_*=e^{-\frac{\sigma_*}{T}-\frac{i\alpha_*}{T}}$. 
When an imaginary chemical potential is present the situation becomes more intriguing. We encounter nontrivial zeros of $D_3(-z_*)$ on the unit circle. A brief exploration in Mathematica revealed two zeroes for $D_3(-z)$ on the unit circle. Remarkably their positions were approximated to high accuracy by rational multiples of $\pi$ as
\begin{align}
\nonumber D_3(-e^{\frac{-i\alpha_*}{T}})=Cl_3(\frac{\alpha_*}{T}\pm\pi)=0\Rightarrow\\ \frac{\alpha_*}{T}\approx \frac{7\pi}{13}\,{\rm or}\,\frac{\alpha_*}{T}=\frac{19\pi}{13} \,\,\,({\rm mod}\,2\pi)\,.
\end{align}
When we use the periodic properties of Clausen functions, we encounter the following relevant results:
\begin{equation}
Cl_3\left(\frac{6\pi}{13}\right)=Cl_3\left(\frac{20\pi}{13}\right)=0.000362159\,.
\end{equation}
and
\begin{equation}
Q_{5f,extr}=\pm i\frac{3T^4}{\pi^2}Cl_4\left(\frac{6\pi}{13}\right)
\end{equation}
since $Cl_4(\pm 6\pi/13)\approx \pm 0.995777$ are the maximum (minimum) values of $D_4(-z)$ on the unit circle. We saw that this pattern generalizes to all dimensions.
Finally, when $\frac{\alpha_*}{T}=\pi$  the gap equation coincides - apart the overall $\sigma_*$ factor - with the corresponding one of the $CP^{N-1}$ that will be given below. The charge is $Q_{5f}=0$ and the system has been bosonized (insulator mode) \cite{Filothodoros:2024}.

The seven-dimensional case shows how our results are generalized to higher dimensions. The gap equation is

\begin{equation}
\frac{4\pi^3}{3T^6}Q_{7f}+15i\left[D_6(-z_*)+\frac{1}{10}\ln^2\!|z_*|D_4(-z_*)
+\frac{1}{120}\ln^4\!|z_*|D_2(-z_*)\right]=0
\end{equation}

Moving on the non zero chemical potential we can look for zeros of the critical gap equation on the unit circle. Again, their positions are remarkably well approximated, better than in $d=5$, by rational multiples of $\pi$ as
\begin{align}
\nonumber D_5(e^{\frac{-i\alpha_*}{T}})=Cl_5(\frac{\alpha_*}{T}\pm \pi)=0\Rightarrow \\
 \frac{\alpha_*}{T}\approx \frac{26\pi}{51}\,{\rm or}\,\frac{76\pi}{51} \,\,\,({\rm mod}\,2\pi)\,.
\end{align}
The relevant result is 
\begin{equation}
Cl_5\left(\frac{25\pi}{51}\right)=Cl_5\left(\frac{77\pi}{51}\right)=0.000129657\,.
\end{equation}
and we found at these points that
\begin{equation}
Q_{7f,extr}=\mp i\frac{45T^6}{4\pi^3 }Cl_6\left(\frac{25\pi}{51}\right)
\end{equation}
since $Cl_6(\pm 25\pi/51)\approx \pm 0.999151$ are  the maximum (minimum) values of $D_6(-z)$ on the unit circle.  

\subsection{\textit{2.3.\,\,The charge $Q_{db}$ of bosons}}
The action of the bosonic theory for general $d$ is from \cite{Filothodoros:2018}
\begin{align}
S_{2b}=\int_0^\frac{1}{T} \!\!\!dx^0 \!\!\int \!\!d^{d-1}x\left[|(\partial_0-i\alpha)\phi^a|^2 +|\partial_i\phi|^2  +i\lambda(\bar{\phi}^a\phi^a-\frac{N}{g_d})+iNQ_{db}\alpha\right]\,,\,\,\,a=1,2,..,N\,,
\end{align}
where the auxiliary scalar field $\lambda$ enforces the constraint $|\phi|^2=N/g_d$ and $Q_{db}$ is the  eigenvalue density of the $N$-normalized $U(1)$ charge density operator $\hat{Q}_{db}=-ig\bar\phi^a\overset{\leftrightarrow}\partial_0\phi^a/N$. In the given model, we have both a global $SU(N)$ symmetry and a global $U(1)$ symmetry. The latter can be easily gauged by introducing a non-propagating abelian gauge field. Upon integrating out the scalar fields, we arrive at the canonical partition function as
\begin{align}
\label{CPPF1}
Z_{2b}(\frac{1}{T},Q_{db})&=\int({\cal D}\alpha)({\cal D}\lambda)e^{-NS_{2b,eff}}\,,\\
\label{CPPF2}
S_{2b,eff}&= iQ_{db}\int_0^\frac{1}{T}\!\!\!dx^0 \!\!\int \!\!d^{d-1}\bar{x}\,\alpha+i\frac{1}{g_d}\int_0^\frac{1}{T}\!\!\!dx^0\!\!\int \!\!d^{d-1}\bar{x}\,\lambda-\rm Tr\ln\left(-(\partial_0-i\alpha)^2-\partial^2+i\lambda\right)_\frac{1}{T}\,.
\end{align}

The $d$-dimensional charge gap equation is 

\begin{equation}
iQ_{db}=-\lim_{\epsilon\rightarrow 0}2T\int\frac{d^{d-1} \bar{p}}{(2\pi)^{d-1}}\sum_{-\infty}^\infty\frac{e^{i\omega_n\epsilon}(\omega_n-\alpha_*)}{\bar{p}^2+(\omega_n-\alpha_*)^2+m_*^2}
\end{equation}

where the bosonic  frequencies are $\omega_n=2\pi nT$ and we have set $i\lambda_*\equiv m_*^2$ in order to facilitate the comparison with the fermionic gap equation (\ref{gap31}) as $\sigma_*$ and $m_*$ have the same dimensions.

The edges of the bosonic thermal windows represents a switch in the statistics where the bosonic becomes fermionic. These are given by roots on the unit circle of $D_{d-2}(z_*)$ and we find for example for $d=5$ and $d=7$
\begin{equation}
\label{D33roots}
D_3(e^{-i \frac{\alpha_*}{T}})=Cl_3(\frac{\alpha_*}{T})=0\,\,\,\Rightarrow\,\,\,\frac{\alpha_*}{T}=\frac{6\pi}{13}\,{\rm or}\, \frac{\alpha_*}{T}=\frac{20\pi}{13}\,,\,\,\,({\rm mod} \,2\pi)\,.
\end{equation}
At the edges of the bosonic thermal windows, the charge becomes extremized. Specifically, it is given by the following expression:
\begin{equation}
\label{qmax}
Q_{5b,extr}=\mp\frac{3iT^4}{4\pi^2}Cl_4\left(\frac{6\pi}{13}\right)=\frac{Q_{5f,extr}}{4}
 \end{equation} 
for $\frac{\alpha_*}{T}=6\pi/3$ and $20\pi/13$ respectively or

\begin{equation}
Q_{7b,extr}=\mp\frac{45iT^6}{32\pi^3}Cl_6\left(\frac{25\pi}{51}\right)=\frac{Q_{7f,extr}}{8}
 \end{equation} 
for $\frac{\alpha_*}{T}=25\pi/51$ and $77\pi/51$.

So if one wants to use a fermionic-bosonic model with equal charges then he has to consider for every boson with charge $Q_b$ a fermion with charge $Q_f=2^{\frac{d-1}{2}}Q_b$, or for every fermion a number of $\frac{d-1}{2}$ bosons, where $d=3,5,7,...$.

\section{3.\,\,Large $N$ fermions and bosons on the lattice}

Consider a system in three dimensions and at finite temperature $T$ with a global $U(1)$ charge operator $\hat{Q}$. Its canonical partition function can be formally calculated as the thermal average over states with fixed $\hat{Q}$ as
\begin{equation}
\label{canPF}
Z_c(\frac{1}{T},B)=Tr\left[\delta(\hat{Q}-B)e^{-\frac{\hat{H}}{T}}\right]
\end{equation}
If the eigenvalues $B$ of $\hat{Q}$ are integers, namely if the system contains elementary excitations, an explicit  representation of (\ref{canPF}) can be written as
\begin{equation}
\label{gcPF}
Z_c(\frac{1}{T},B)=\int_{0}^{2\pi}\!\frac{d\theta}{2\pi}\,e^{i\theta B}\,\rm Tr\left[e^{\frac{\hat{H}}{T}-i\theta\hat{Q}}\right]=\int_{0}^{2\pi}\!\frac{d\theta}{2\pi}\,e^{i\theta B}\,Z_{gc}(\frac{1}{T},i\mu=i\theta T),
\end{equation}
where $Z_{gc}(T,i\mu)$ is the grand canonical partition function with imaginary chemical potential $i\mu$ \cite{D'Elia, Lombardo, Alford, Kapustin}. 

  In the simple systems we are interested in one expects that 
\begin{equation}
\label{Zgcperiodic}
Z_{gc}(\frac{1}{T},i(\mu +2\pi k T))= Z_{gc}(\frac{1}{T},i\mu)\,,\,\,\,k\in \mathbb{Z}
\end{equation}
One then notices with Bloch's theorem as follows \cite{Filothodoros:2023}. In quantum system, taken here to be $1d$ for clarity,  in a periodic potential with period $a$ the energy eigenstates are the Bloch waves
\begin{equation}
\label{Blochwaves}
\psi_k(x) =e^{ikx}u(x)\,,\,\,\,u(x+a)=u(x)
\end{equation}
where $k$ is the lattice momentum vector \cite{Blochoverlaps}. The transition amplitude between two Bloch waves with different lattice momenta is
\begin{equation}
\label{transition}
\langle \psi_{k_1}|\psi_{k_2}\rangle =\int_0^a dx e^{i(k_2-k_1)x}|u(x)|^2
\end{equation}

In spherical coordinates we may write
\begin{equation}
\label{sphercoo}
\langle \psi_{k_1}|\psi_{k_2}\rangle =\frac{1}{2\pi}\int_0^{2\pi} d\theta e^{i(k_2-k_1)\theta}|u(\theta)|^2
\end{equation}

Notice that $B$ may be thought of as a charge transfer 'momentum'. According to the theoretical background of the Hubbard model, this charge transfer 'momentum' is a potential term that allows the fermions to travel on lattice sites. This is like a term that describes the transition between conducting and insulating systems. When $B$ is $0$ the Hubbard-like lattice is an insulator and when $B$ is max the lattice is a conductor. 

A suitable expression of a $1d$ Bloch-wave is of the form:

\begin{equation}
\psi_{k}(\theta)=e^{ik\theta} u(\theta)
\end{equation}
where $u$ function has the periodicity of the lattice $2\pi$ since $u(\theta)=u(\theta+2\pi)$, like $Z_{gc}(\beta,i\mu=i\theta T)$ has the periodicity of the chemical potential. The fermionic states in a Hubbard model is described by the Wannier functions that defined by

\begin{equation}
\Phi_X(x)=\frac{1}{\sqrt{N}}\sum_{k}e^{-ikX}\psi_{k}(x)
\end{equation}
where $X$ is a Bravais lattice vector in cartesian coordinates for a $1d$ quantum system \cite{Ambrosetti, Marzari, Bermudez, Liao}.
According to the replacement rule the sum can be written as an integral, since $N$ is very large, like

\begin{equation}
\sum_{k}=\frac{N}{V}\int_{BZ}dk
\end{equation}
where $V$ is the volume of the $BZ$ Brillouin zone.

Interestingly the Bloch functions can be written in terms of Wannier functions like
\begin{equation}
\psi_k(x)=\frac{1}{\sqrt{N}}\sum_{X}e^{ikX}\Phi_{X}(x)
\end{equation}

So, in spherical coordinates the $\psi$ Bloch function may be written as
\begin{equation}
\psi_{k}(\theta)=\frac{1}{2\pi}\int_0^{2\pi}d\theta e^{ik\Theta}\Phi_{\Theta}(\theta)
\end{equation}
where $\Theta$ is a Bravais lattice vector in spherical coordinates. So the picture is that the canonical partition function of the Gross Neveu model is like a Bloch state that varies continuously with $k$ (or $B$) and the grand canonical partition function is like a Wannier state at imaginary chemical potential $i\Theta$. The grand canonical partition function is a localized system with periodic boundary conditions arising from the imaginary chemical potential. This periodicity is an interesting quasi-periodicity since we have the appearance of the Bloch-Wigner-Ramakrishnan polylogarithm function $D_m(z)$. It is interesting that this function exhibits a very interesting behaviour regarding periodicity. For $m=even$, the function is not periodic since its sum involves alternating terms, and the series does not repeat in a regular pattern. For $m=odd$, the function has a more intricate structure. While the individual terms in the series are not periodic, the presence of the logarithmic term $\log^m|x|$ introduces a quasi-periodicity. Specifically the function satisfies the functional 
\begin{equation}
D_m(1/x)=(-1)^{m-1}D_m(x)
\end{equation} 

Then we have two parts in partition function $Z_c$. The charge part with $even$ index of $D_m(z)$ function and the $\rm Tr\log$ and the grand canonical part with $odd$ index of $D_m(z)$. The first part has no periodicity but the second part has a quasi-periodicity of the hypothetical quasi-crystal of the band theory. When we take the points of $D_{m}(z)$ on the unit circle we have the Clausen function which is periodic with period $2\pi$ and the matching between Wannier function and the grand canonical partition function is obvious. The grand canonical parts of $Z_{1f}$ and $Z_{2b}$ are an orthogonal set of Wannier functions where

\begin{equation}
\int_0^{2\pi} Z_{1f_{gc}}Z_{2b_{gc}}=0
\end{equation}

So the transition amplitude on the unit circle between the two Bloch states of the Gross-Neveu and $CP^{N-1}$ models is from \cite{Filothodoros:2018}:

\begin{equation}
\label{pfdualityd0}
Z^{d}_{fb}(\frac{\alpha_*}{T})\Bigl|_{\sigma_*=0}\equiv\left[Z_{1f}^{(d)}(\frac{\alpha_*}{T}+\pi)Z_{2b}^{(d)}(\frac{\alpha_*}{T})\right]\Bigl|_{\sigma_*=0}=e^{4\pi N\frac{V_{d-1}}{S_d T^{1-d}}Cl_{d-1}(\frac{\alpha_*}{T})}
\end{equation}
The advantage here is that the divergent terms from the bosonic partition function's exponential come with the opposite sign compared to the fermionic partition function so their contribution is zero. At zero temperature $T\rightarrow 0$ the ratio $V_{d-1}/S_{d}T^{1-d}\rightarrow 1$ (i.e. we can think of $V_{d-1}$ as the surface of a very large sphere). Then the Clausen's functions with even index at $d\rightarrow\infty$ limits are $\sin(\frac{\alpha_*}{T})$. Therefore we write
\begin{equation}
\label{pfsusylimit}
\lim_{d\rightarrow\infty} Z^{d}_{fb}(\frac{\alpha_*}{T})\Bigl|_{\sigma_*=0}=e^{4\pi N\sin(\frac{\alpha_*}{T})}\,.
\end{equation}
 
Suppose that we have a large number of $N$ fermions (blue) and $N$ bosons (red) on Euclidean space, where the charge of every fermion is $2^{\frac{d-1}{2}}$ times the bosonic charge. Here $d$ plays the role of the order of the family of lattices covering the Euclidean space. We also suppose that the large Euclidean surface is covered by regular hexagons with the help of the imaginary chemical potential mentioned before and let us assume that the fermions and bosons are on the vertices of these lattices.
\begin{figure}[!htb]
\centering
\includegraphics[scale=0.44]{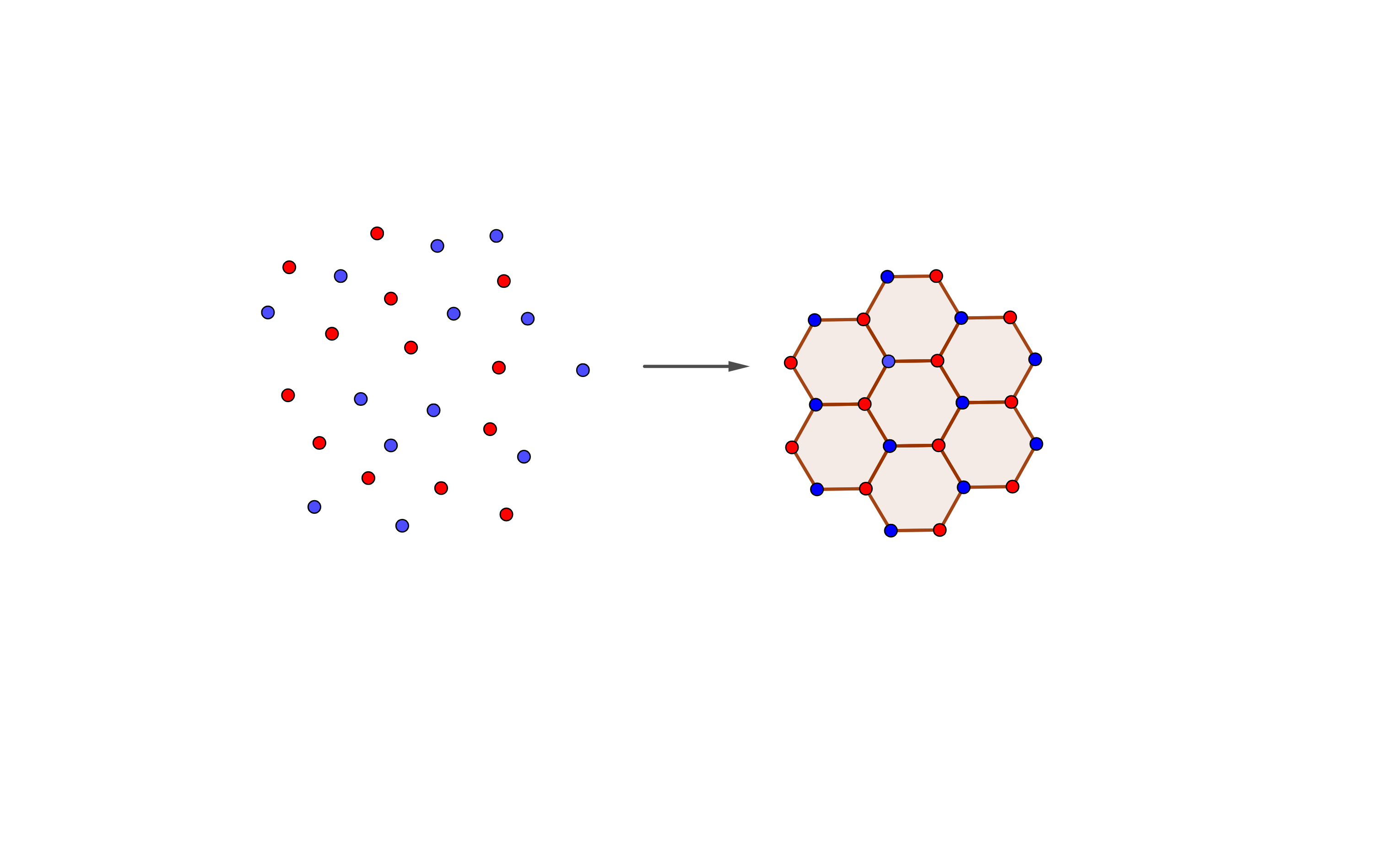}
\caption{Fermions and bosons on Euclidean surface}
\end{figure}

Let's see for example the case of the regular hexagon and its "perturbed" relatives of the hexagons conjecture. Based on the hypothesis we had previously, we can, in line with the generalization of the thermodynamic study of Gross-Neveu and CP$^{N-1}$ theories, give a more general hypothesis on the observation that the inner angles of a regular hexagon are exactly $2\pi/3$ and there is a lattice point at $2\pi/3$ on the unit circle so the maximization of Clausen function which is equal to the imaginary part of $Li_d(-z)$, where $d=2,4,6,...$, on the unit circle for specific angles $7\pi/13$,$26\pi/51$,$103\pi/205$ and so on. These values come from an analytic formula for the approximate positions of the zeros of all $D_{2n-1}(z)$, $n=1,2,..$  functions on the unit circle. We obtain:
\begin{equation}
	D_{2n-1}(e^{-i\frac{\alpha_*}{T}})\equiv Cl_{2n-1}(\frac{\alpha_*}{T})=0\Leftrightarrow \frac{\alpha_*}{T}\approx \theta_n,2\pi -\theta_n\,({\rm mod}\,2\pi)\
\end{equation}
where
\begin{equation}
\label{ThetaN}
	\theta_n=\frac{\pi}{2}\left(1-\frac{5}{4^{n+1}-(-1)^{n+1}}\right)\
\end{equation}
for $n=1,2,3,..$. There is an interesting approximation of these zeros in \cite{Etienne}.

The particles are the lattice points on Euclidean space of irregular hexagons. At a large $d$ limit the conjecture ends up to the square lattice construction. Somehow as dimension increases the 6th and 5th sides disappear and the particles are getting closer to make some pairs. We may say that these pairs are similar to Cooper pairs. The conjecture turns  to be as follows:

{\centering
	\begin{tabular}{|c|c|c|c|c|c|c|}
		\hline 
		\multicolumn{7}{|c|}{\textbf{Table 1. Positions of the lattice points on the unit circle for hexagonal model}} \\ 
		\hline 
		Order $d$ & Lattice point $1$ & Lattice point $2$ & Lattice point $3$ & Lattice point $4$ & Lattice point $5$ & Lattice point $6$ \\
		\hline 
		3 & $\pi/3$ & $2\pi/3$ & $\pi$ & $4\pi/3$ & $5\pi/3$ & $2\pi$\\  
		\hline 
		5 & $6\pi/13$ & $7\pi/13$ & $\pi$ & $19\pi/26$ & $20\pi/26$ & $2\pi$\\ 
		\hline 
		7 & $25\pi/51$ & $26\pi/51$ & $\pi$ & $76\pi/51$ & $77\pi/51$ & $2\pi$\\ 
		\hline 
		9 & $102\pi/205$ & $103\pi/205$ & $\pi$ & $307\pi/205$ & $308\pi/205$ & $2\pi$\\ 
		\hline 
		... & ... & ... & ... & ... & ... & ... \\ 
		\hline 
		$\infty$ & $\pi/2$ & $\pi/2$ & $\pi$ & $3\pi/2$ & $3\pi/2$ &$2\pi$ \\ 
		\hline 
	\end{tabular}\par}

One may see the transformation of a hexagonal Hubbard lattice to a square one and compare with the phase transitions of the Gross-Neveu model and $CP^{N-1}$ model at imaginary chemical potential. The picture is like the one that follows:

''Studying the correspondence of thermal windows with the transformation of a hexagonal Hubbard lattice to square lattice we have fermions at high temperature (edges of a fermionic thermal window) and bosons at the beginning of their fermionization area (edges of a bosonic thermal window). The thermal windows close for both of them and we have fermions with the same charge that getting closer to each other, with a strong interaction (clockwise for fermions to $\pi/2$ and anti-clockwise for fermions to $3\pi/2$, anti-clockwise for bosons to $\pi/2$ and clockwise for bosons to $3\pi/2$). The picture is like the Hubbard model of electrons in a periodic potential which predicts the theory of conductivity. Fermions from hexagons (order $3$ lattice) are on lattice points $2\pi/3$, $4\pi/3$ and $0$ and the fermionized bosons are on $\pi/3$, $5\pi/3$ and $\pi$  (the bosons are in the beginning of their "fermionization" area for the $CP^{N-1}$ model) and moving to make a pair with fermions. We see that the chemical potential is moving them closer and closer until they coincide on angle $\pi/2$ on the unit circle where $D_{even}$ function has its maximum value for the Gross-Neveu model ($D_{\infty}(\pi/2)=Cl_{\infty}(\pi/2)=sin(\pi/2)=1$). At order $\infty$ of the model we have a bound of two fermions with total charge $2Q_f$, at $\pi/2$ and $3\pi/2$. We must not forget that while in the Hubbard model the fermions have the same charge, in the two models of fermions and bosons the fermions and bosons have charges connected by the relationship $Q_f=2^{\frac{d-1}{2}}Q_b$. This picture could correspond to the creation of Cooper pairs of conductivity.''

\subsection{\textit{3.1.\,\,Generalized Bragg Law}}

Imagine a beam of particles with a wavelength similar to the spacing between atoms in a crystal.
When this beam encounters a lattice plane within the crystal, it gets scattered.
The incident and reflected waves from the lattice plane remain in phase if the difference in their path lengths which determines the constructive interference, is given by the formula: $n\lambda=2d\sin\theta$, where $\theta$ is the scattering angle and $d$ is the interplanar distance.

Suppose now that a particle travels in a crystal lattice with lattice spacing $d$, with momentum $q_i$ and collides with a lattice point with an exchange of momentum $q=q_i-q_f$, where $q_f$ is the the momentum of the reflected particle. Since a wave has momentum $q=\hbar k$ ($k$ is the wave number) in reciprocal space momentum transfer is given by $Q=k_i-k_f$ ($k$ is a vectorial quantity). Then we are able to calculate a reciprocal lattice vector $G=Q$ with the relation to the lattice spacing as $G=2\pi/d$. Consequently the total transfer momentum from N particles of a beam scattered by a crystal lattice is:

\begin{equation}
Q_{total}=4\pi N sin\theta
\end{equation}

For a perturbed version of Bragg Law we may define a new function $\Delta$ where
\begin{equation}
	\Delta(\theta)\rightarrow \left( sin\theta+\sum_{k=2}^{\infty}\frac{sink\theta}{k^m}\right)
\end{equation}
for arbitrary even $m=2,4,...,\infty$. This $\Delta$ function is Clausen function $Cl_m(\theta)$.	
So the Bragg Law of diffraction of a lattice with lattice space $d$ may have a generalized version:
\begin{equation}
	n\lambda=2d\Delta(\theta)\rightarrow n\lambda=2d\left( sin\theta+\sum_{k=2}^{\infty}\frac{sink\theta}{k^m}\right).
\end{equation}

The ratio $sin\theta/\Delta(\theta)$ for  $m\rightarrow \infty$ is $1$, follows from the  observation that this limit is actually well defined
\begin{equation}
\label{ClausenInfinity}\lim_{m\rightarrow\infty}Cl_{m}(\theta)=\sin\theta\,
\end{equation}
 where $m=2n, n=1,2,3,4...$..

\begin{figure}[!htb]
 \centering
 \includegraphics[scale=0.52]{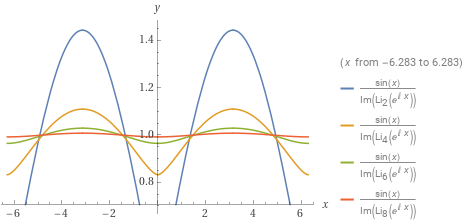}
\caption{The ratio $sin\theta/\Delta(\theta)$ for $\theta\in[-2\pi,2\pi]$}
\end{figure}
 
The sum in the parenthesis is the Clausen function so:
\begin{equation}
	n\lambda=2d Cl_m(\theta)	
\end{equation}
The generalized transferred momentum is 
\begin{equation}
	Q_{gen}=G
\end{equation}
where $G$ is the reciprocal lattice vector with $G=2\pi/d$. For $n=1$, $\lambda=1$ and $N$ beam particles we have:
\begin{equation}
	Q_{gen}=4\pi N Cl_m(\theta)
\label{Bragg}
\end{equation}

So, (\ref{pfdualityd0}) may be equivalent to the calculation of the overlap between two Bloch wavefunctions that differ by lattice momentum $Q_{gen}$ when one is the incident and the other the reflected wave from a lattice point.

\section{4.\,\,Generalized thermal windows and lattice points}

Let's see in detail Figure 3. Interestingly there is an equivalent picture of the thermal windows in the Gross-Neveu and $CP^{N-1}$ models at finite T and imaginary chemical potential. In the Gross-Neveu model and its change in the statistics we have temperature (the chemical potential) that "creates" the thermal windows, where inside them we have a chiral symmetry breaking in order to make some fermion condensates. Outside these windows we have Fermi statistics. On the other hand the imaginary chemical potential "creates" the fermionization thermal windows for bosons where it is always possible to tune the coupling to a special value, such that the thermal mass vanishes. Outside these windows we have Bose statistics. This picture is reminiscent to the behaviour of the Gross-Neveu model where for a nonzero temperature it was possible to tune the coupling in order to restore parity. With the conjecture of the thermal windows we have the equivalent picture of the region inside the hexagon.  Later, we will observe that the boundaries of the thermal windows align with the positions of lattice points on the hexagonal lattice. At these points, $D_2(-z*)$ has its maximum value.
	
	From \cite{Filothodoros:2018} the generalized thermal windows are:\\
	
	{\centering
		\begin{tabular}{|c|c|c|}
			\hline 
			\multicolumn{3}{|c|}{\textbf{Table 2. Generalized Thermal Windows for the $GN$ model}} \\ 
			\hline 
			Dimensions & Closing T & Opening T \\ 
                        \hline
                        3& $\frac{3\alpha_*}{4\pi}$&$\frac{3\alpha_*}{2\pi}$\\
			\hline 
			5 & $\frac{13\alpha_*}{19\pi}$ & $\frac{13\alpha_*}{7\pi}$ \\ 
			\hline 
			7 & $\frac{51\alpha_*}{76\pi}$ & $\frac{51\alpha_*}{26\pi}$ \\ 
			\hline 
			9 & $\frac{205\alpha_*}{307\pi}$ & $\frac{205\alpha_*}{103\pi}$ \\ 
			\hline 
			11 & $\frac{819\alpha_*}{1228\pi}$ & $\frac{819\alpha_*}{410\pi}$ \\ 
			\hline 
		\end{tabular}\par} 

Let's focus on the case of the $3d$ theory thermal window. The borders for $\alpha_*$ are $2\pi T/3$ and $4\pi T/3$. These are the points where the $D_2(-z*)$ takes its maximum value (imaginary part) on the unit circle. On the unit circle $D_2(-z*)=Cl_2(\pi-\frac{\alpha_*}{T})$. The generalized thermal windows for the $CP^{N-1}$ model are:\\

 {\centering
		\begin{tabular}{|c|c|c|}
			\hline 
			\multicolumn{3}{|c|}{\textbf{Table 3. Generalized Thermal Windows for the $CP^{N-1}$ model}} \\ 
			\hline 
			Dimensions & Closing T & Opening T \\ 
                        \hline
                        3& $\frac{3\alpha_*}{5\pi}$&$\frac{3\alpha_*}{\pi}$\\
			\hline 
			5 & $\frac{13\alpha_*}{20\pi}$ & $\frac{13\alpha_*}{6\pi}$ \\ 
			\hline 
			7 & $\frac{51\alpha_*}{77\pi}$ & $\frac{51\alpha_*}{25\pi}$ \\ 
			\hline 
			9 & $\frac{205\alpha_*}{308\pi}$ & $\frac{205\alpha_*}{102\pi}$ \\ 
			\hline 
			11 & $\frac{819\alpha_*}{1229\pi}$ & $\frac{819\alpha_*}{409\pi}$ \\ 
			\hline 
		\end{tabular}\par} 

Let's focus again on the case of the $3d$ theory thermal window. The borders for $\alpha_*$ are $\pi T/3$ and $5\pi T/3$. These are the points where the $D_2(z*)$ takes its maximum value (imaginary part) on the unit circle. On the unit circle $D_2(z*)=Cl_2(\frac{\alpha_*}{T})$. The interesting aspect involves placing all these points on the unit circle which is circumscribed about a regular hexagon as we have explained before. So the lattice points of the "supersymmetric" model are lying on the unit circle at $\alpha=0, \pi/3, 2\pi/3, \pi, 4\pi/3, 5\pi/3$. The "supersymmetric" model when the regular hexagon turns to square has a superconducting picture arising from the identities of the a bosonic and fermionic models at imaginary chemical potential.

\begin{figure}
 \centering
\includegraphics[scale=0.44]{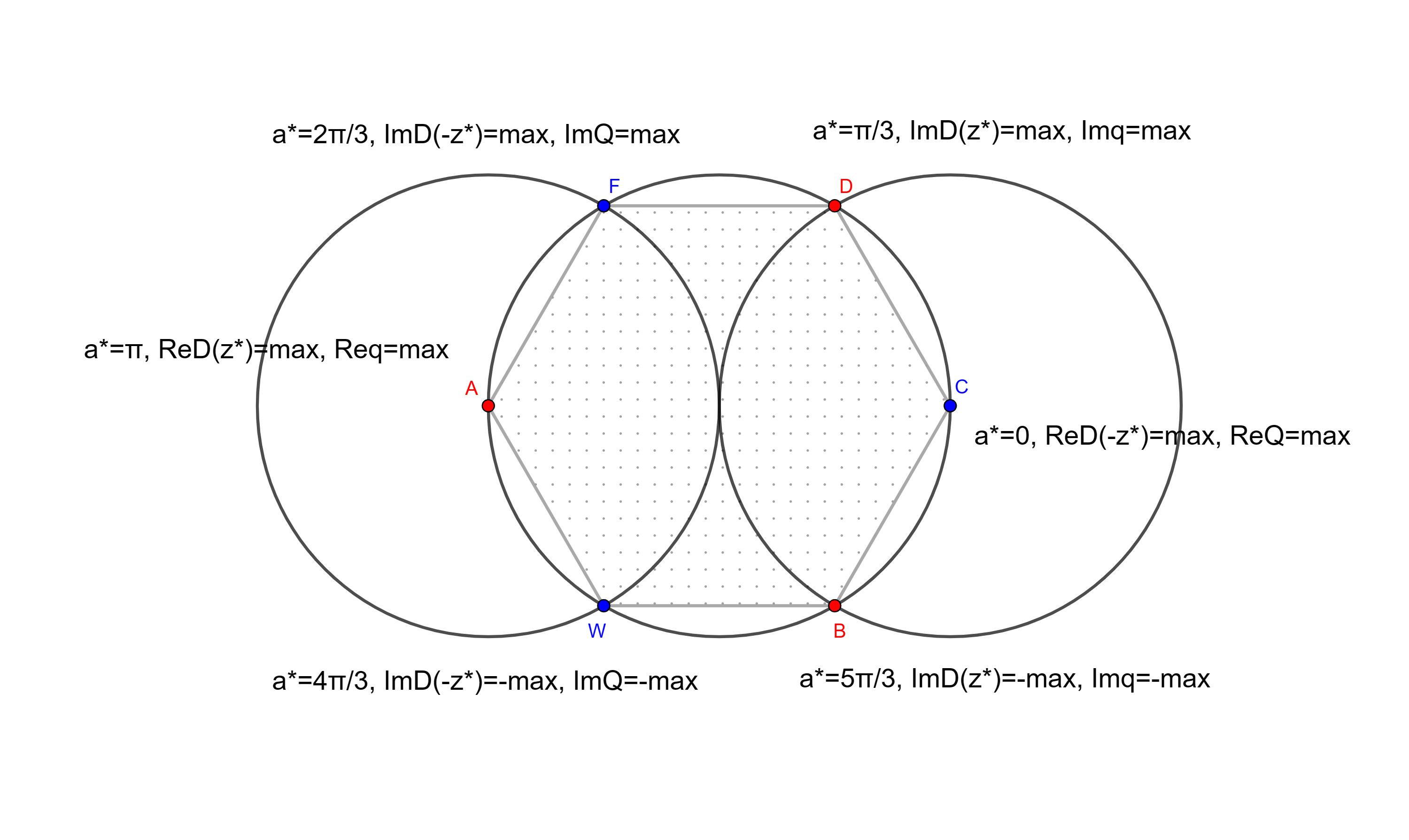}
\caption{GN and CP thermal windows on lattice}
\end{figure}

\begin{figure}
 \centering
 \includegraphics[scale=0.68]{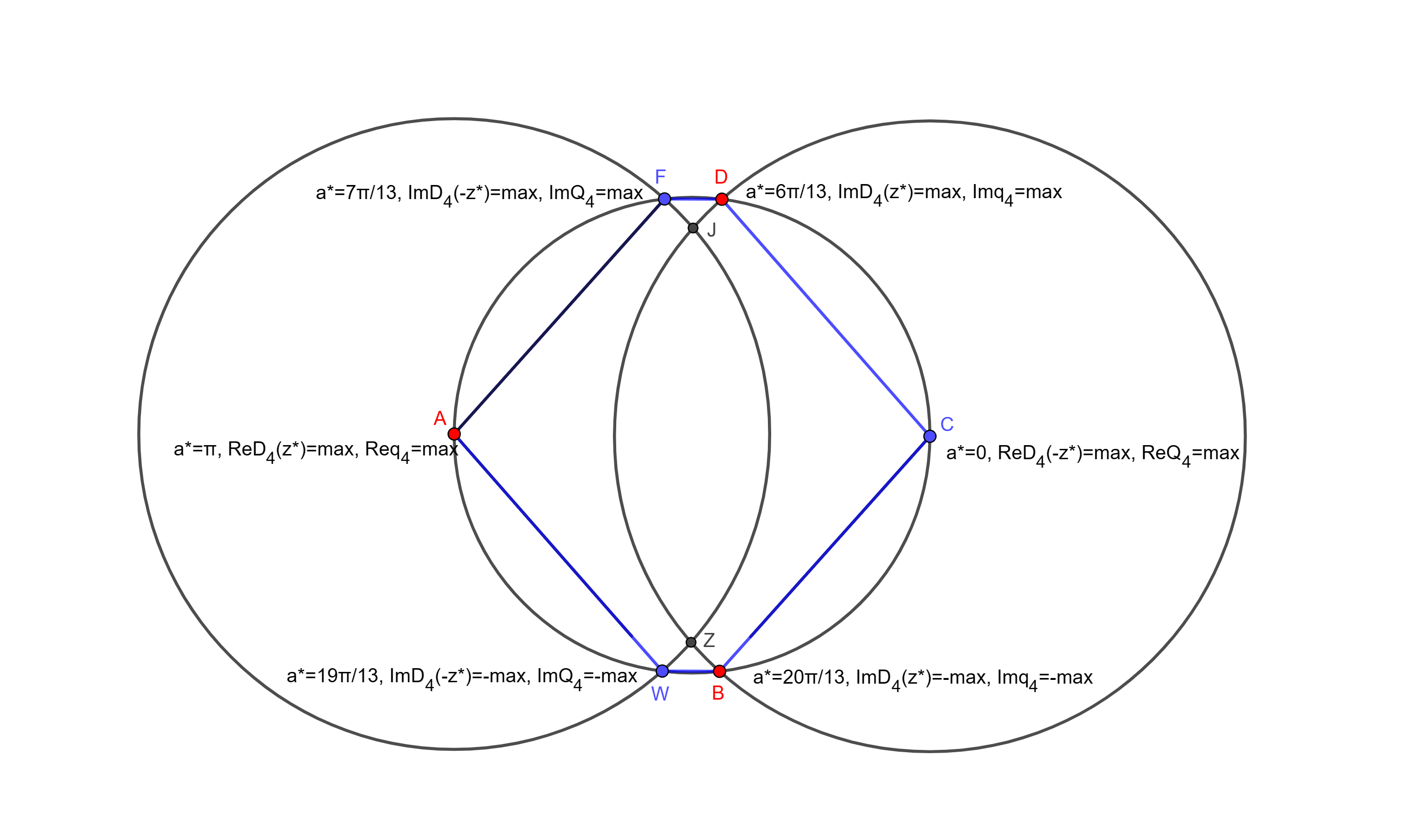}
\caption{GN and CP thermal windows on perturbed lattice}
\end{figure}

\begin{figure}[!htb]
 \centering
\includegraphics[scale=0.42]{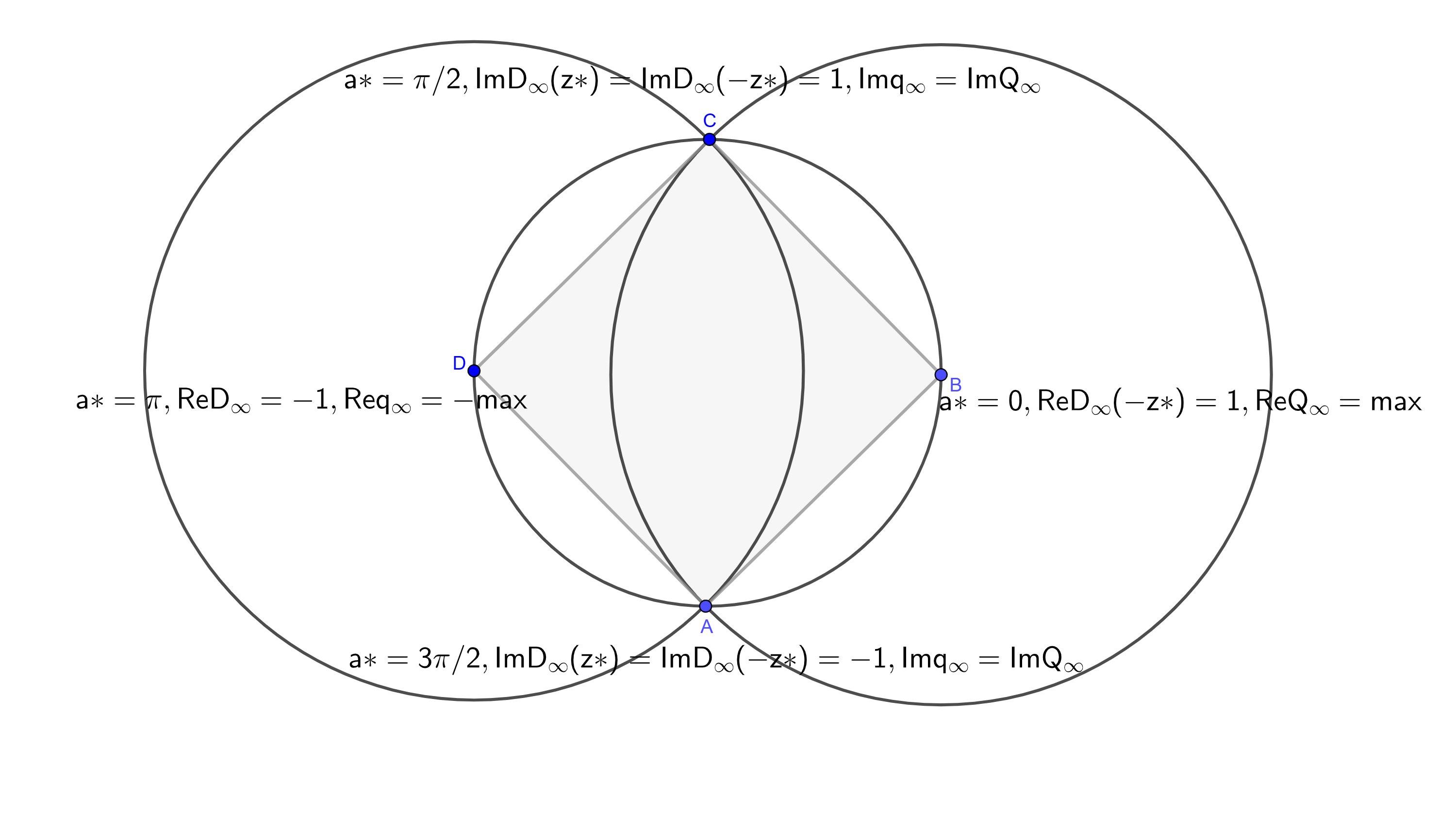}
\caption{GN and CP thermal windows on square lattice}
\end{figure}
	
\section{5.\,\,Summary and discussion}
The main message of this work is that a fermionic and a bosonic theory at imaginary chemical potential for arbitrary odd dimensions, have a phase structure which is connected to a hexagonal Hubbard model of fermions. 	Particularly, the fermion-boson duality at imaginary chemical potential enables us to relate the thermal windows of the fermionic $U(N)$ Gross-Neveu model and the bosonic CP$^{N-1}$ model at finite temperature to lattice transformations of hexagonal Hubbard-like lattices to square lattice. Theoretical studies have predicted that certain systems can undergo a transition from a Hubbard hexagonal lattice to a square lattice. Since our calculations have unveiled the relevance of the canonical partition functions of the models to Bloch states, we think that our results offer a new window into the utility of the Gross-Neveu and $CP^{N-1}$ models as a tool with which the creation of a Cooper-pair can be studied. The overlap between two Bloch wavefunctions that differ by lattice momentum $Q$ may be equivalent to the calculation of the total charge of a Cooper-pair of fermions or the momentum transfer that one Bloch wave gives to a lattice when reflected by a lattice point.



  
  
\section*{Statement of Competing Interests}

The author report that he does not have 
any competing interests associated with this research work.

\appendix

\section{The Bloch-Wigner-Ramakrishnan functions $D_m(z)$}
From the usual analytic continuation of the polylogarithms
\begin{equation}
Li_m(z)=\sum_{n=1}^\infty\frac{z^n}{n^m}\,,\,\,\,z\in{\mathbb C}\setminus [1,\infty)\,,\,\,\,m=1,2,3,..\,.
\end{equation}
one can define the following Bloch-Wigner-Ramakrishnan functions \cite{Zagier1,Zagier2} as
\begin{equation}
\label{Ds}
D_m(z)=\Re\left(i^{m+1}\left[\sum_{k=1}^m\frac{(-\ln|z|)^{m-k}}{(m-k)!}Li_k(z)-\frac{(-\ln|z|)^m}{2m!}\right]\right)
\end{equation}
These are real functions of complex variable, analytic in ${\mathbb C}\setminus \{0,1\}$.
In the text we utilized the following properties of $D_m(z)$'s.
\begin{align}
\label{Dd1}
D_m(1/z)&=(-1)^{m-1}D_m(z)\\
\label{Dd2}
 \frac{\partial}{\partial z}D_m(z)&=\frac{i}{2z}\left(D_{m-1}(z)+\frac{i}{2}\frac{(-i\ln|z|)^{m-1}}{(m-1)!}\frac{1+z}{1-z}\right)
 \end{align}
On the unit circle  we have
\begin{align}
\label{Dodd}
&D_{2n-1}(e^{-i\theta})=(-1)^n\Re[Li_{2n-1}(e^{-i\theta})]=(-1)^n Cl_{2n-1}(\theta)\,,\\
\label{Deven}
&D_{2n}(e^{-i\theta})=(-1)^{n+1}\Im[Li_{2n}(e^{-i\theta})]=(-1)^{n}Cl_{2n}(\theta)\,
\end{align}
for $n=1,2,3,..$. 
The Clausen functions $Cl_m(\theta)$ are defined as
\begin{equation}
\label{Clausen2}
Cl_{2n-1}(\theta)\equiv\sum_{k=1}^{\infty}\frac{\cos k\theta}{k^{2n-1}}\,,\,\,\,Cl_{2n}(\theta)\equiv \sum_{k=1}^{\infty}\frac{\sin k\theta}{k^{2n}}\,,\,\,\,n=1,2,..
\end{equation}

\end{document}